\documentclass[aps,twocolumn,superscriptaddress,showpacs]{revtex4}
\usepackage{graphicx}
\begin{document}
\draft
\title{\bf  The Union Jack model: a quantum spin model with frustration on
the square lattice}
\author{ A. Collins, J. McEvoy, D. Robinson, C. J. Hamer and Zheng Weihong}
\affiliation{School of Physics, The University of New South Wales,
  Sydney, NSW 2052, Australia}
\date{\today}
\begin{abstract}
A new quantum spin model with frustration, the `Union Jack' model
on the square lattice, is analyzed using spin-wave theory. For
small values of the frustrating coupling $\alpha$, the system is
N{\' e}el ordered as usual, while for large $\alpha$ the
frustration is found to induce a canted phase. The possibility of
an intermediate spin-liquid phase is discussed.
\end{abstract}
\pacs{PACS Indices: 05.30.-d,75.10.-b,75.10.Jm,75.30.Ds,75.30.Kz \\
\\  \\
(Submitted to  Phys. Rev. B) }
\maketitle
\newpage

\section{INTRODUCTION}
\label{sec1}

\begin{figure}
 \includegraphics[width=0.9\linewidth]{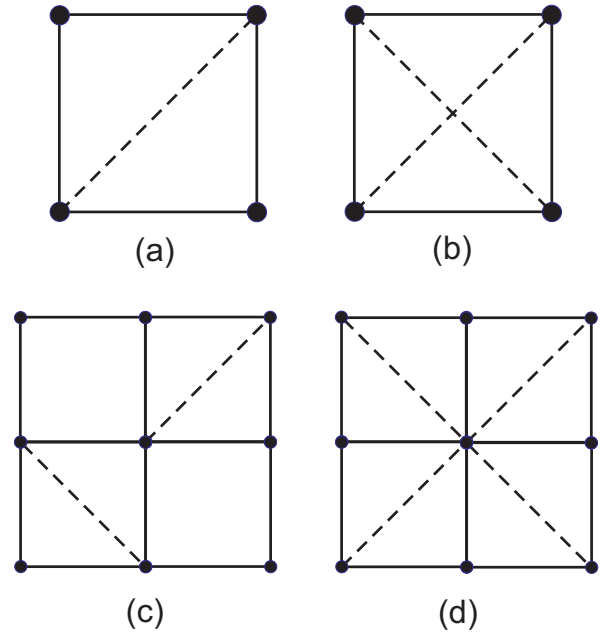}
 \caption{ Lattice spin models with frustration. Solid lines represent
nearest-neighbour antiferromagnetic interactions $J_1$; dashed lines
represent next-nearest-neighbour interactions $J_2$.
 Case: a) anisotropic triangular lattice; b) $J_1-J_2$ model; c)
 Shastry-Sutherland model; d) Union Jack model.}
 \label{fig1}
\end{figure}

Frustrated lattice spin models in two dimensions have attracted
much discussion in recent years. They exhibit new and interesting
phase structures and phase transitions; in particular, they may
develop `spin liquid' states, without long-range order
\cite{anderson1987}. It is also believed that frustrating
interactions may play a role in the high-temperature
superconducting cuprate materials.  Primary examples are the
anisotropic triangular lattice Heisenberg antiferromagnet (Fig.
\ref{fig1}a), the square lattice $J_1-J_2$ model (Fig.
\ref{fig1}b), and the Shastry-Sutherland model (Fig. \ref{fig1}c).
In this paper we discuss another member of this group, the `Union
Jack' model (Fig. \ref{fig1}d), which is another square lattice
Heisenberg antiferromagnet with frustration, and might be expected
to display some interesting properties.

The spin-1/2 $J_1-J_2$ model on the square lattice
\cite{chandra1988} involves antiferromagnetic Heisenberg spin
interactions with coupling $J_1$ between nearest neighbours, and
coupling $J_2$ between diagonal next-nearest neighbours, as
illustrated in Figure \ref{fig1}b). The Union Jack model has $J_2$
interactions on only half the diagonal bonds, in the pattern shown
in Figure \ref{fig1}d), so that the lattice consists of two
different site types, A and B, and the unit cell is 2x2 sites.
Both models will exhibit quantum phase transitions as the coupling
ratio $\alpha=J_2/J_1$ is varied.

\begin{figure}
 \includegraphics[width=0.9\linewidth]{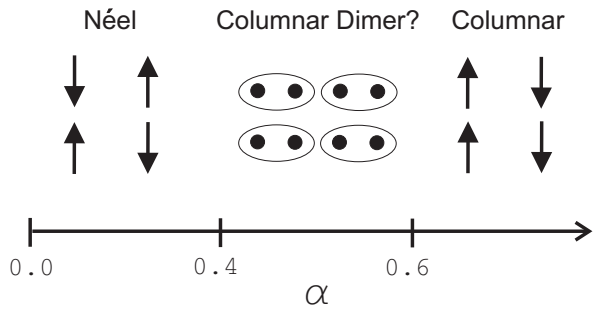}
 \caption{ Phase diagram of the $J_1-J_2$ model.}
 \label{fig2}
\end{figure}

In the $J_1-J_2$ model at small $\alpha$, the $J_1$ coupling is dominant and
produces antiferromagnetic N{\' e}el ordering of the spins. At large
$\alpha$, the $J_2$ interaction is dominant, and produces antiferromagnetic
ordering on the two diagonal sublattices; and then the effect of the
$J_1$ interaction is to align the two sublattices to form a columnar
ordered state as illustrated in Figure \ref{fig2}, an example of the
`order by disorder' phenomenon. Numerical
investigations \cite{dagotto1989,schulz1996,kotov1999,gelfand1989,
singh1999,sushkov2001,capriotti2000,jongh2000} have shown that the boundaries of these two
phases lie at $\alpha \simeq 0.38$ and $\alpha \simeq 0.60$, respectively.

The nature of the intermediate phase or phases remains controversial.
Monte Carlo simulations are hampered by the `minus sign' problem; exact
diagonalizations are limited to small lattices; and series expansions
are based on some particular ordered reference state, and are only valid
within a single phase. It is generally believed that the intermediate
phase is gapped and shows no long-range magnetic order. Field theory
approaches \cite{read1991,kotov1999}
and dimer series expansions \cite{kotov1999,gelfand1989,singh1999,sushkov2001}
seem to indicate a columnar
dimerized state in the intermediate region, with spontaneous breaking of
translational symmetry, as illustrated in Figure \ref{fig2}. Capriotti, Sorella
and coworkers, on the other hand,
have suggested  a homogeneous spin-liquid
`plaquette RVB' state \cite{capriotti2000,sorella2003}, and have found
that exact
diagonalization up to 6x6 sites shows no strong evidence of dimerization
\cite{capriotti2003}. Another Monte Carlo study has suggested a columnar dimer
state with
plaquette-type modulation \cite{jongh2000}. Sushkov {\it et al.} \cite{sushkov2001} have
even suggested that there may be three different phases in the
intermediate region: reading from left to right, a N{\' e}el state
with columnar dimerization; a columnar dimerized spin liquid, and a
columnar dimerized spin liquid with plaquette-type modulation.

Several discussions have centered on the Lieb-Schulz-Mattis
theorem in higher dimensions
\cite{yamanaka1997,oshikawa2000,hastings2004}, which shows that
for a spin system with half-integer spin per unit cell, there is
an excitation energy gap behaving like $1/L$, where $L$ is the
linear size of the system. Takano {\it et al.} \cite{takano2003}
have argued that a uniform RVB state without gapless singlet
excitations is excluded by the theorem, and that the true ground
state is a plaquette state with spontaneously broken translation
invariance and fourfold degeneracy. Later arguments
\cite{sorella2003,misguich2002,hastings2004} have refuted this,
however, and shown that the theorem may be satisfied if the
translation symmetry remains unbroken, but the ground state has a
fourfold topological degeneracy instead, as in a simple dimer
model.

\begin{figure}
 \includegraphics[width=0.7\linewidth]{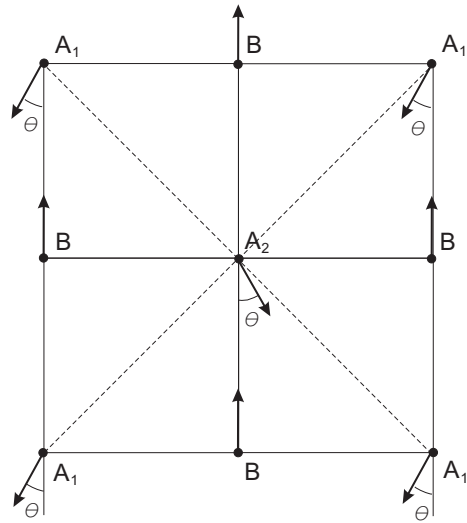}
 \caption{Spin configurations in the canted phase. }
 \label{fig3}
\end{figure}

The Hamiltonian for the Union Jack model is
\begin{equation}
H = J_1 \sum_{<nn>} {\bf S_i \cdot S_j} + J_2 \sum_{A:<nnn>} {\bf S_i
\cdot S_j}
\label{eq1}
\end{equation}
where the $J_1$ and $J_2$ interactions connect sites as shown in Figure
\ref{fig1}d). A classical variational analysis shows that for $\alpha =J_2/J_1 <
0.5$, the ground state is the N{\' e}el state, as in the $J_1-J_2$
model. For $\alpha  > 0.5$, the ground state is the canted ferrimagnetic state
shown in Figure \ref{fig3}, where the spins on the A sublattice are
canted at an angle $\theta$ to those on the B sublattice, and $2\theta$
to their neighbours on the A sublattice. The energy of this state is
\begin{equation}
E_0 = NS^2[\alpha \cos 2\theta - 2\cos \theta]
\label{eq2}
\end{equation}
where we have set $J_1=1$, $S$ is the total spin per site, and $N$ is
the number of sites. This energy is minimized when
\begin{equation}
\sin \theta (2\alpha \cos \theta -1) = 0.
\label{eq3}
\end{equation}
For $\alpha  < 0.5$, the lowest energy corresponds to $\sin \theta = 0$, i.e.
the simple N{\' e}el state. For $\alpha  > 0.5$, the lowest energy solution is
\begin{equation}
\cos \theta = \frac{1}{2\alpha },
\label{eq4}
\end{equation}
corresponding to the canted state. In the limit $\alpha \rightarrow
\infty$, the angle $\theta \rightarrow \pi/2$: the spins on the A sublattice are
N{\' e}el ordered, as expected, and the spins on the A and B sublattices
are at right angles to each other.

In the canted phase, there is a net magnetization per site in the
z direction,
\begin{equation}
M_z = \frac{S}{2}(1-\cos \theta) = \frac{S}{2}(1-\frac{1}{2\alpha })
\label{eq5}
\end{equation}
and a staggered magnetization per site
\begin{equation}
M_s = \frac{S}{2}(1+\cos \theta) = \frac{S}{2}(1+\frac{1}{2\alpha }).
\label{eq6}
\end{equation}

The model thus demonstrates a new mechanism for causing ferrimagnetism.
Ferrimagnetism usually arises when the individual ionic spins are
different on different sublattices. Here the spins have the same
magnitude on each sublattice, and the interactions are
antiferromagnetic, but the frustration between them produces a nonzero
overall magnetic moment.

In the neighbourhood of the transition point, $\alpha_c = 0.5$, we find
for the ground state energy per bond $\epsilon_0 = E_0/2N$ in zero
magnetic field :

\begin{eqnarray}
\epsilon_0 &  \sim & -\frac{3S^2}{8}, \hspace{5mm} \alpha \rightarrow
\alpha_c \\
\frac{d\epsilon_0}{d\alpha} &  \sim & S^2/2, \hspace{5mm} \alpha \rightarrow
\alpha_c \\
\frac{d^2\epsilon_0}{d\alpha^2} &  \sim & \left\{ \begin{array}{c}
 0, \hspace{5mm} \alpha \rightarrow \alpha_c- \\
 -4S^2, \hspace{5mm} \alpha \rightarrow \alpha_c+
\end{array} \right.
\end{eqnarray}
while for the net magnetization in the $z$ direction,
\begin{equation}
M_z \sim S(\alpha-\alpha_c), \hspace{5mm} \alpha \rightarrow \alpha_c.
\end{equation}
Thus the transition is not quite the simple first-order transition found
classically for other frustrated models. There is no ``latent heat",
i.e. no discontinuity in $d\epsilon_0/d\alpha$, although there is a
finite discontinuity in the second derivative. The magnetization $M_z$
vanishes linearly at $\alpha_c$, and the energy gap is zero on both
sides of the transition, corresponding to spontaneous breaking of the
spin rotation symmetry. Translation symmetry on the lattice is also
broken on both sides of the transition. The question then arises,
whether quantum fluctuations will change the character of the
transition.

In the remainder of this paper, we shall discuss how this picture is
modified when one goes beyond the classical analysis to a spin-wave
treatment. In Section \ref{sec2} we shall present a modified spin-wave
treatment of the N{\' e}el phase to second order;
and in
Section \ref{sec3} we shall give a spin-wave treatment of the
canted phase to first order. The results are discussed in Section
\ref{sec4}. Our conclusions are summarized in
Section \ref{sec5}.

\section{Spin-wave theory for the N{\' e}el phase.}
\label{sec2}

\subsection{Formulation}

A review of spin-wave approaches to the $J_1-J_2$ model has been given
by Gochev \cite{gochev1994}. Chandra and Doucot \cite{chandra1988} gave
the first conventional spin-wave treatment. The results of this approach
were found to become unstable as the frustration $\alpha$ increases, due
to strong interactions between the spin wave bosons. The interactions
involve a quadratic term
\begin{equation}
\tilde{H}_2 = \sum_{\bf k} Q_{\bf k}(\alpha_{\bf k}\beta_{\bf k}+
\alpha^{\dagger}_{\bf k} \beta^{\dagger}_{\bf k})
\label{eq7}
\end{equation}
which is comparable with the zeroth-order Hamiltonian for $S \sim 1$ and
$\alpha \sim 1$. Thus one needs to go to a modified treatment in which
this quadratic term is absent \cite{xu1990,nishimori1990}: this
requirement turns out to be equivalent to Takahashi's modified spin-wave
theory \cite{takahashi1987}. A very
similar treatment can be applied to the Union Jack model. We will follow
Gochev's notation \cite{gochev1994} as far as possible.

Let us rewrite the Hamiltonian (\ref{eq1}), adding in a staggered
magnetic field $h$:
\begin{equation}
H = \sum_{<lm>}{\bf S_l \cdot S_m} + \alpha \sum_{<l_1l_2>} {\bf S_{l_1}
\cdot S_{l_2}} +h(\sum_l S^z_l - \sum_m S^z_m)
\label{eq8}
\end{equation}
where we have divided the lattice into even (A) and odd (B) sublattices,
denoted by indices $l$ and $m$ respectively, and set $J_1 = 1$.

Introduce creation and annihilation operators for the ``spin deviation"
on the two sublattices by means of a Dyson-Maleev transformation:

\begin{eqnarray}
S^z_l & = & S - a^{\dagger}_la_l \nonumber \\
S^+_l & =  & (2S)^{1/2}a_l - (2S)^{-1/2}a^{\dagger}_la_la_l \nonumber \\
 S^-_l & = &
(2S)^{1/2}a^{\dagger}_l  \nonumber \\
 S^z_m & = & b^{\dagger}_mb_m - S \nonumber \\
S^+_m & = & (2S)^{1/2}b^{\dagger}_m
-(2S)^{-1/2}b^{\dagger}_mb^{\dagger}_mb_m \nonumber \\
 S^-_m & = & (2S)^{1/2}b_m
\label{eq9}
\end{eqnarray}

and then perform a Fourier transform

\begin{eqnarray}
a_{\bf k} & = & \left(\frac{2}{N}\right)^{1/2} \sum_{\bf l} e^{i{\bf k \cdot l}}a_{\bf l};
 \nonumber \\
b_{\bf k} & = & \left(\frac{2}{N}\right)^{1/2} \sum_{\bf m} e^{-i{\bf k \cdot m}}b_{\bf m}
\label{eq10}
\end{eqnarray}
to give the Hamiltonian in the form
\begin{widetext}
\begin{eqnarray}
H & = & -2S^2N(1-\frac{\alpha}{2}-\frac{h}{2S})
+ 4S\{(1-\alpha-\frac{h}{4S})\sum_{\bf k} a^{\dagger}_{\bf k}a_{\bf k}
+(1-\frac{h}{4S}) \sum_{\bf k} b^{\dagger}_{\bf k}b_{\bf k} +
  \sum_{\bf k} [\gamma_{\bf k}(a_{\bf k}b_{\bf k} + a^{\dagger}_{\bf
k}b^{\dagger}_{\bf k}) + \alpha \eta_{\bf k}a^{\dagger}_{\bf k}a_{\bf
k}]\} \nonumber \\
 & & -\frac{4}{N}\sum_{\bf 1234}\delta_{{\bf 1-2-3+4}}[2\gamma_{\bf
3-4}a^{\dagger}_{\bf 1}a_{\bf 2}b^{\dagger}_{\bf 3}b_{\bf 4} +
\gamma_{\bf 4}a^{\dagger}_{\bf 1}a_{\bf 2}a_{\bf 3}b_{\bf 4} +
\gamma_{\bf 1}a^{\dagger}_{\bf 1}b^{\dagger}_{\bf 2}b^{\dagger}_{\bf
3}b_{\bf 4}] \nonumber \\
 & &
  + \frac{4\alpha}{N} \sum_{\bf 1234} \delta_{\bf 1+2-3-4}
a^{\dagger}_{\bf 1}a^{\dagger}_{\bf 2}a_{\bf 3}a_{\bf 4}(\eta_{\bf 2-4}
- \frac{1}{2}(\eta_{\bf 1} + \eta_{\bf 2}))
\label{eq11}
\end{eqnarray}
\end{widetext}
where we have used the shorthand notation ${\bf 1 \cdots 4}$ for
momenta ${\bf k_1 \cdots k_4}$, and $\gamma_{\bf k},\eta_{\bf k}$
are the structure factors for the full lattice and the A
sublattice respectively:
\begin{eqnarray}
\gamma_{\bf k} & = & \cos \frac{k_x}{2} \cos \frac{k_y}{2} \nonumber \\
\eta_{\bf k} & = & \frac{1}{2}(\cos k_x + \cos k_y),
\label{eq12}
\end{eqnarray}
if we set the spacing of each sublattice equal to 1, and $k_x,k_y$ are
the components of momentum along the diagonal axes of the two
sublattices.

The Hamiltonian can now be diagonalized up to second order by a
Bogoliubov transformation:
\begin{equation}
a_{\bf k} = u_{\bf k}\alpha_{\bf k} - v_{\bf k}\beta^{\dagger}_{\bf k},
\ \
b_{\bf k} = u_{\bf k}\beta_{\bf k}-v_{\bf k}\alpha^{\dagger}_{\bf k}
\label{eq13}
\end{equation}
where
\begin{equation}
u_{\bf k}^2 - v_{\bf k}^2 = 1.
\label{eq14}
\end{equation}

After normal ordering the transformed Hamiltonian, the condition that
off-diagonal quadratic terms vanish turns out to be:
\begin{eqnarray}
Q_{\bf k} &  = & 4S[\gamma_{\bf k}(u^2_{\bf k} + v^2_{\bf k}) - 2u_{\bf
k}v_{\bf k}(1-\frac{h}{4S}-\frac{\alpha}{2}(1-\eta_{\bf k}))] \nonumber
\\
 & & +4(R_1-R_2)[\gamma_{\bf k}(u^2_{\bf k} + v^2_{\bf k})-2u_{\bf
k}v_{\bf k}]  \nonumber \\
 & & - 4\alpha u_{\bf k}v_{\bf k}(R_2-R_3)(1-\eta_{\bf k})
\nonumber \\
 & = & 0
\label{eq15}
\end{eqnarray}
where the lattice sums $R_i$ are
\begin{equation}
R_1 = \frac{2}{N}\sum_{\bf k} \gamma_{\bf k}u_{\bf k}v_{\bf k}, \ R_2 =
\frac{2}{N}\sum_{\bf k} v^2_{\bf k}, \ R_3 = \frac{2}{N} \sum_{\bf k} \eta_{\bf
k}v^2_{\bf k}
\label{eq16}
\end{equation}

A solution to equation (\ref{eq15}) can easily be found (for $h \leq 0$):
\begin{equation}
u_{\bf k} = \left[\frac{1+\epsilon_{\bf k}}{2\epsilon_{\bf k}}\right]^{1/2}, \
v_{\bf k} = {\rm sgn}(\gamma_{\bf k})\left[\frac{1-\epsilon_{\bf k}}{2\epsilon_{\bf
k}}\right]^{1/2} \nonumber
\label{eq17}
\end{equation}
\begin{eqnarray}
\epsilon_{\bf k} & = & \left(1-\frac{\gamma_{\bf k}^2}{f_{\bf k}^2}\right)^{1/2},
\nonumber \\
f_{\bf k} & = & 1 - h\sigma - \rho \frac{\alpha}{2} (1-\eta_{\bf k})
\label{eq18}
\end{eqnarray}
where
\begin{eqnarray}
\sigma & = & \frac{1}{4(S+R_1-R_2)}
\nonumber \\
\rho & = & \frac{S+R_3-R_2}{S+R_1-R_2}
\label{eq19}
\end{eqnarray}
(the parameter $\rho$ is Gochev's $\bar{\alpha}$). Then one merely
has to find self-consistent solutions for the two parameters $\sigma$
and $\rho$, given by equations (\ref{eq19}).

We can now write the Hamiltonian as
\begin{equation}
H_{DM} = W_0 + H_0 + V_{DM}.
\label{eq20}
\end{equation}
The constant term is
\begin{eqnarray}
W_0 & = & 2N\epsilon_0
  =  2N[-(S+R_1-R_2)^2 \nonumber \\
 & &  +\frac{\alpha}{2}(S+R_3-R_2)^2 +
\frac{h}{2}(S-R_2)]
\label{eq21}
\end{eqnarray}
(note that there is a misprint in Gochev \cite{gochev1994} at this
point). The quadratic part $H_0$ is diagonal:
\begin{equation}
H_0 = \sum_{\bf k} (E^{\alpha}_{\bf k}\alpha^{\dagger}_{\bf k}\alpha_{\bf k} +
E^{\beta}_{\bf k}\beta^{\dagger}_{\bf k}\beta_{\bf k})
\label{eq22}
\end{equation}
where the $\alpha$ and $\beta$ bosons have different spin-wave energies
in this case:
\begin{eqnarray}
E^{\alpha}_{\bf k} & = & 4S[u^2_{\bf k}(1-\alpha(1-\eta_{\bf k}) -
\frac{h}{4S}) + v^2_{\bf k}(1-\frac{h}{4S}) \nonumber \\
 & &  -2u_{\bf k}v_{\bf
k}\gamma_{\bf k}]
 +4(R_1-R_2)(u^2_{\bf k} + v^2_{\bf k} - 2u_{\bf k}v_{\bf
k}\gamma_{\bf k})  \nonumber \\
 & & + 4\alpha u^2_{\bf k}(R_2-R_3)(1-\eta_{\bf k})
\label{eq23}
\end{eqnarray}
and $E^{\beta}_{\bf k}$ is a similar expression with $u_{\bf k}$ and
$v_{\bf k}$ interchanged.

The normal-ordered quartic interaction operator $V_{DM}$ is
\begin{widetext}
\begin{eqnarray}
V_{DM} & = & -\frac{2}{N}\sum_{\bf 1234} \delta_{\bf 1+2-3-4}
[\Phi^{(1)}\alpha_{\bf 1}\alpha_{\bf 2}\beta_{\bf 3}\beta_{\bf 4} +
\Phi^{(2)}\alpha^{\dagger}_{\bf 3}\alpha^{\dagger}_{\bf
4}\beta^{\dagger}_{\bf 1}\beta^{\dagger}_{\bf 2}
 -2\Phi^{(3)}\alpha^{\dagger}_{\bf 3}\beta_{\bf 4}\alpha_{\bf
1}\alpha_{\bf 2}-2\Phi^{(4)}\alpha^{\dagger}_{\bf 4}\beta^{\dagger}_{\bf
1}\beta^{\dagger}_{\bf 2}\beta_{\bf 3} \nonumber \\
 & & -2\Phi^{(5)}\beta^{\dagger}_{\bf
2}\beta_{\bf 3}\beta_{\bf 4}\alpha_{\bf 1}
  -2\Phi^{(6)}\alpha^{\dagger}_{\bf 3}\alpha^{\dagger}_{\bf
4}\beta^{\dagger}_{\bf 1}\alpha_{\bf 2}+\Phi^{(7)}\beta^{\dagger}_{\bf
1}\beta^{\dagger}_{\bf 2}\beta_{\bf 3}\beta_{\bf 4}
+\Phi^{(8)}\alpha^{\dagger}_{\bf 3}\alpha^{\dagger}_{\bf 4}\alpha_{\bf
1}\alpha_{\bf 2}+4\Phi^{(9)}\alpha^{\dagger}_{\bf 4}\beta^{\dagger}_{\bf
1}\alpha_{\bf 2}\beta_{\bf 3}].
\label{eq24}
\end{eqnarray}
\end{widetext}
Explicit expressions for the vertex functions $\Phi^{(i)}({\bf 1234})$
of the N{\' e}el phase are given in Appendix A.

\subsection{Higher-order corrections}

To order 1 in a $1/S$ expansion, the ground-state energy per bond is
$\epsilon_0$, as given by equation (\ref{eq21}).
The staggered magnetization per site is
\begin{equation}
M_s = 2\frac{\partial\epsilon_0}{\partial h}|_{h=0} = S - R_2,
\label{eq25}
\end{equation}
and the spin-wave energy is that given by equation (\ref{eq23}).

\begin{figure}
 \includegraphics[width=0.9\linewidth]{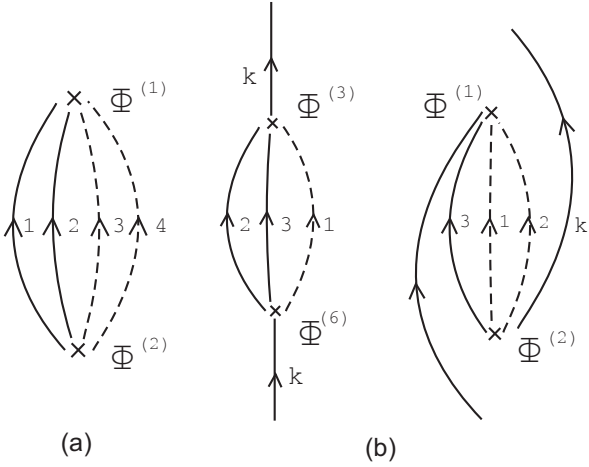}
 \caption{Perturbation diagrams in the N{\' e}el phase. Solid lines
correspond to $\alpha$ bosons, dashed lines to $\beta$ bosons. a)
Leading correction to the ground-state energy. b) Leading corrections to
the spin-wave energy $E^{\alpha}_{\bf k}$.}
 \label{fig4}
\end{figure}

We can now use perturbation theory to calculate the next-order ($1/S$)
corrections to these results. The leading correction to the ground-state
energy corresponds to Figure \ref{fig4}a), and is given by
\begin{equation}
\frac{\Delta E_0}{2N} = -\left(\frac{2}{N}\right)^3\sum_{\bf 1234}\delta_{\bf
1+2-3-4}\frac{\Phi^{(1)}({\bf 1234})\Phi^{(2)}({\bf
3412})}{(E^{\alpha}_{\bf 1} + E^{\alpha}_{\bf 2} + E^{\beta}_{\bf 3} + E^{\beta}_{\bf
4})},
\label{eq26}
\end{equation}
while the corrections to the spin-wave energies $E^{\alpha}_{\bf k}$ and
$E^{\beta}_{\bf k}$ (Fig. \ref{fig4}b) are
\begin{widetext}
\begin{eqnarray}
\Delta E^{\alpha}_{\bf k}  & = &  -8\left(\frac{2}{N}\right)^2\sum_{\bf 123}\delta_{\bf
2+3-1-k}\frac{\Phi^{(6)}({\bf 1k23})\Phi^{(3)}({\bf
23k1})}{(E^{\beta}_{\bf 1} + E^{\alpha}_{\bf 2} + E^{\alpha}_{\bf 3} - E^{\alpha}_{\bf
k})}
 -8\left(\frac{2}{N}\right)^2\sum_{\bf 123}\delta_{\
1+2-3-k}\frac{\Phi^{(2)}({\bf 123k})\Phi^{(1)}({\bf
k321})}{(E^{\beta}_{\bf 1} + E^{\beta}_{\bf 2} + E^{\alpha}_{\bf 3} + E^{\alpha}_{\bf
k})}, \nonumber \\
\Delta E^{\beta}_{\bf k} &  = &  -8\left(\frac{2}{N}\right)^2\sum_{\bf 123}\delta_{\bf
2+3-1-k}\frac{\Phi^{(5)}({\bf 1k23})\Phi^{(4)}({\bf
23k1})}{(E^{\alpha}_{\bf 1} + E^{\beta}_{\bf 2} + E^{\beta}_{\bf 3} - E^{\beta}_{\bf
k})}
 -8\left(\frac{2}{N}\right)^2\sum_{\bf 123}\delta_{\
1+2-3-k}\frac{\Phi^{(1)}({\bf 123k})\Phi^{(2)}({\bf
k321})}{(E^{\alpha}_{\bf 1} + E^{\alpha}_{\bf 2} + E^{\beta}_{\bf 3} + E^{\beta}_{\bf
k})}
\label{eq27}
\end{eqnarray}
\end{widetext}

The corrections to the staggered magnetization can also be represented
in terms of perturbation diagrams as follows. We have

\begin{figure}
 \includegraphics[width=0.9\linewidth]{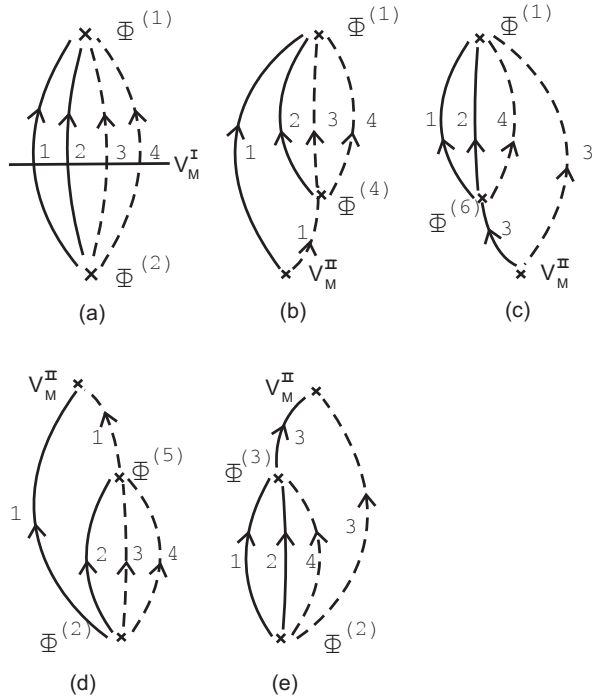}
 \caption{ Perturbation diagrams contributing to the N{\' e}el
magnetization at order $1/S$.}
 \label{fig5}
\end{figure}

\begin{eqnarray}
M_s & = & \frac{1}{N}<\psi_0|\sum_lS^z_l - \sum_m S^z_m|\psi_0> \\
 & = & <\psi_0| S -R_2 - V_M^{I} -V_M^{II}|\psi_0>
\label{eq27a}
\end{eqnarray}
where

\begin{eqnarray}
V_M^I & = & \sum_{\bf k}(u_{\bf k}^2 + v_{\bf k}^2) (\alpha^{\dagger}_{\bf k}\alpha_{\bf k} +
\beta^{\dagger}_{\bf k}\beta_{\bf k}) \\
V_M^{II} & = & \sum_{\bf k} 2u_{\bf k}v_{\bf k} (\alpha^{\dagger}_{\bf k}\beta^{\dagger}_{\bf k} +
\alpha_{\bf k}\beta_{\bf k})
\label{eq27b}
\end{eqnarray}
Hence the leading corrections to $M_s$ of $O(1/S)$ correspond to the
diagrams shown in Figure \ref{fig5}, and are given by

\begin{widetext}
\begin{eqnarray}
\Delta M_s & = & -\frac{16}{N^3}\sum_{\bf 1234} \delta_{\bf
1+2-3-4}\frac{\Phi^{(1)}({\bf 1234})\Phi^{(2)}({\bf
3412})}{(E_{\bf 1}^{\alpha}+E_{\bf 2}^{\alpha}+E_{\bf 3}^{\beta}+E_{\bf 4}^{\beta})^2}\left[\frac{1}
{\epsilon_{\bf 1}}+\frac{1}{\epsilon_{\bf 2}}+\frac{1}{\epsilon_{\bf 3}}+\frac{1}{\epsilon_{\bf
4}}\right]
\\
 & & -\frac{32}{N^3}\sum_{\bf 1234}\delta_{\bf 1+2-3-4}\frac{F({\bf
1234})}{(E_{\bf 1}^{\alpha}+E_{\bf 2}^{\alpha}+E_{\bf 3}^{\beta}+E_{\bf 4}^{\beta})}
\label{eq27c}
\end{eqnarray}
where
\begin{equation}
F({\bf 1234}) = \Phi^{(1)}({\bf 1234})\left[\frac{\gamma_{\bf
1}\Phi^{(4)}({\bf 3412})}{f_{\bf 1}\epsilon_{\bf 1}(E_{\bf
1}^{\alpha}+E_{\bf 1}^{\beta})}+\frac{\gamma_{\bf
3}\Phi^{(6)}({\bf 4312})}{f_{\bf 3}\epsilon_{\bf 3}(E_{\bf
3}^{\alpha}+E_{\bf 3}^{\beta})}\right] + \Phi^{(2)}({\bf
3412})\left[\frac{\gamma_{\bf 1}\Phi^{(5)}({\bf 2134})}{f_{\bf
1}\epsilon_{\bf 1}(E_{\bf 1}^{\alpha}+E_{\bf
1}^{\beta})}+\frac{\gamma_{\bf 3}\Phi^{(3)}({\bf 1234})}{f_{\bf
3}\epsilon_{\bf 3}(E_{\bf 3}^{\alpha}+E_{\bf 3}^{\beta})}\right].
\label{eq27d}
\end{equation}
\end{widetext}

The results of these calculations are
discussed in Section \ref{sec4}.

\section{Spin-wave theory for the canted phase.}
\label{sec3}

In this section we shall present a modified spin wave treatment of the
canted phase. The formalism turns out to be somewhat more
complicated in this case, and we shall only keep terms down to O(S) in
the energy. We write the Hamiltonian as

\begin{eqnarray}
H & = & \sum_{<nn>} {\bf S_i \cdot S_j} +\alpha \sum_{A:<nnn>} {\bf S_i
\cdot S_j} \nonumber \\
 & &  + h_1 \sum_{B:i} S^z_i +h_2 \sum_{A:i} S^z_i,
\label{eq29}
\end{eqnarray}
quantizing the spins with respect to the configuration shown in
Figure 3, so that the $z$ axis on the B sublattice points upwards,
and the $z$ axes on the sublattices A${_1}$ and A${_2}$ are canted
at angle $\theta$ to the downwards direction as shown, where
$\theta$ is a parameter to be determined. The $y$ axes are taken
to lie perpendicular to the paper in each case. In terms of spin
components, we then have:
\begin{widetext}
\begin{eqnarray}
H & = & \sum_{B:{\bf n,\mu}} \left[-(S^z_{B:{\bf n}}S^z_{A:{\bf
n+\mu}}+S^x_{B:{\bf n}}S^x_{A:{\bf n+\mu}})\cos \theta + \eta_{{\bf
n\mu}}(S^z_{B:{\bf n}}S^x_{A:{\bf n+\mu}}-S^x_{B:{\bf n}}S^z_{A:{\bf
n+\mu}})\sin \theta + S^y_{B:{\bf n}}S^y_{A:{\bf n+\mu}} \right] \nonumber
\\
 & & + \alpha \sum_{A1:{\bf n,\mu '}} \left[(S^z_{A1:{\bf n}}S^z_{A2:{\bf
n+\mu '}} + S^x_{A1:{\bf n}}S^x_{A2:{\bf n+\mu '}})\cos 2\theta +
S^y_{A1:{\bf n}}S^y_{A2:{\bf n+\mu '}} +(S^z_{A1:{\bf n}}S^x_{A2:{\bf n+\mu
'}} -S^x_{A1:{\bf n}}S^z_{A2:{\bf n+\mu '}})\sin 2\theta \right]
\nonumber \\
 & &  +
h_1\sum_{B:{\bf n}} S^z_{B:{\bf n}} +h_2\sum_{A:{\bf n}} S^z_{A:{\bf n}}
\label{eq30}
\end{eqnarray}
\end{widetext}
where the direction vectors are $\{{\bf \mu}\} = \pm {\bf i},\pm {\bf j}$; $\{{\bf \mu '}\} = \pm
({\bf i \pm j})$ and the phase factor $\eta_{{\bf n\mu}} = \pm (-1)^{n_y}$
for ${\bf \mu}$ equals $\pm {\bf i}$ or $\pm {\bf j}$, respectively.

Now we introduce boson creation and annihilation operators for the spin deviation
on the two sublattices by means of the Dyson-Maleev transformation:

\begin{eqnarray}
 S^z_{A:{\bf m}} & = & S - a^{\dagger}_{\bf m}a_{\bf m}  \nonumber \\
 S^+_{A:{\bf m}} & = & (2S)^{1/2}a_{\bf m} \nonumber \\
S^-_{A:{\bf m}} & = & (2S)^{1/2}a^{\dagger}_{\bf m}
-(2S)^{-1/2}a^{\dagger}_{\bf m}a^{\dagger}_{\bf m}a_{\bf m} \nonumber \\
S^z_{B:{\bf n}} & = & S - b^{\dagger}_{\bf n}b_{\bf n} \nonumber \\
S^+_{B:{\bf n}} & =  & (2S)^{1/2}b_{\bf n} - (2S)^{-1/2}b^{\dagger}_{\bf n}b_{\bf n}b_{\bf n} \nonumber \\
 S^-_{B:{\bf n}} & = &
(2S)^{1/2}b^{\dagger}_{\bf n}
\label{eq31}
\end{eqnarray}
and perform a Fourier transform

\begin{eqnarray}
a_{\bf k} & = & \left(\frac{2}{N}\right)^{1/2} \sum_{\bf n} e^{i{\bf k \cdot n}}a_{\bf n};
 \nonumber \\
b_{\bf k} & = & \left(\frac{2}{N}\right)^{1/2} \sum_{\bf m} e^{i{\bf k \cdot m}}b_{\bf m}
\label{eq32}
\end{eqnarray}
to obtain
\begin{widetext}
\begin{eqnarray}
H & = & -NS^2\left(2\cos \theta -\alpha \cos 2\theta -\frac{(h_1+h_2)}{2S}\right)
  +2(NS^3)^{1/2}(\sin \theta -\alpha \sin 2\theta)(a_{\pi,\pi} +
a^{\dagger}_{\pi,\pi}) \nonumber \\
 & & +2S[(2\cos \theta -\frac{h_1}{2S}) \sum_{{\bf k}}b^{\dagger}_{{\bf k}}b_{{\bf
k}} +(2(\cos \theta -\alpha \cos 2\theta)-\frac{h_2}{2S})\sum_{{\bf k}} a^{\dagger}_{{\bf
k}}a_{{\bf k}}-(1+\cos \theta)\sum_{{\bf k}}\gamma_{{\bf k}}(b_{{\bf k}}a_{-{\bf
k}} +b^{\dagger}_{{\bf k}}a^{\dagger}_{-{\bf k}}) \nonumber \\
 & &
  +(1-\cos \theta)\sum_{{\bf k}}\gamma_{{\bf k}}(b_{{\bf
k}}a^{\dagger}_{{\bf k}} + b^{\dagger}_{{\bf k}}a_{{\bf
k}})+\frac{\alpha}{2}\sum_{{\bf k}}\eta_{{\bf k}}[2(1+\cos 2\theta)a^{\dagger}_{{\bf
k}}a_{{\bf k}}-(1-\cos 2\theta)(a_{{\bf k}}a_{-{\bf
k}}+a^{\dagger}_{{\bf k}}a^{\dagger}_{-{\bf k}})]]
\label{eq33}
\end{eqnarray}
\end{widetext}
to order $S$,
where as usual
\begin{equation}
\gamma_{{\bf k}} = \frac{1}{4}\sum_{{\bf \mu}}e^{i{\bf k \cdot \mu}},
\hspace{5mm} \eta_{{\bf k}} = \frac{1}{4}\sum_{{\bf \mu}'}e^{i{\bf k
\cdot \mu}'}.
\label{eq34}
\end{equation}

The term of order $S^2$ is simply the classical energy of this
configuration, equation (\ref{eq2}). The term of order
$S^{3/2}$, proportional to $(b_{\pi,\pi}+b^{\dagger}_{\pi,\pi})$, would
indicate that we have not chosen the optimum reference state, and should
be set to zero: this gives a condition on the angle $\theta$ identical
to the classical condition, equation (\ref{eq3}). At higher orders, this
condition would give a `renormalised' value for the angle $\theta$.

The
term of order $S$ is quadratic in fermion operators, and can be
diagonalized by a Bogoliubov transformation. In the sector involving
momenta $(\pm {\bf k})$, the quadratic part of the Hamiltonian is

\begin{equation}
H_{{\bf k}}  =  2S[\left( \begin{array}{cccc}
a^{\dagger}_{{\bf k}} & b_{-{\bf k}} & b^{\dagger}_{{\bf k}} & a_{-{\bf
k}} \end{array} \right)
{\hat H}_{{\bf k}} \left( \begin{array}{c}
a_{{\bf k}} \\
b^{\dagger}_{-{\bf k}} \\
b_{{\bf k}} \\
a^{\dagger}_{-{\bf k}} \end{array} \right) + N_{{\bf k}}]
\label{eq35}
\end{equation}
where to leading order the matrix ${\hat H}_{{\bf k}}$ has elements $h_{ij}$ given
by
\begin{equation}
h_{11} = h_{44} = 2(\cos \theta - \alpha \cos 2\theta) + \alpha (1+\cos
2\theta)\eta_{{\bf k}} -\frac{h_2}{2S}; \nonumber
\end{equation}
\begin{equation}
h_{22} = h_{33} = 2\cos \theta - \frac{h_1}{2S}; \nonumber
\end{equation}
\begin{equation}
h_{12} = h_{21} = h_{34} = h_{43} = -(1+\cos \theta)\gamma_{{\bf k}};
\nonumber
\end{equation}
\begin{equation}
h_{13} = h_{31} = h_{24} = h_{42} = (1-\cos \theta)\gamma_{{\bf k}};
\nonumber
\end{equation}
\begin{equation}
h_{23} = h_{32} = 0.
\label{eq36a}
\end{equation}
\begin{equation}
h_{14} = h_{41} = -\alpha(1-\cos 2\theta)\eta_{{\bf k}};
\label{eq36}
\end{equation}
Note that the matrix ${\hat H}_{{\bf k}}$ is symmetric about {\it both}
diagonals. The normal-ordering correction to $O(S)$ is
\begin{equation}
N_{{\bf k}} = - h_{11} -h_{22}
\label{eq37}
\end{equation}

Thus the procedure here involves a 4x4 matrix diagonalization, rather
than 2x2. We want a transformation which preserves commutation
relations, and the symmetries of the problem, i.e. we need to find a {\it
symplectic} transformation:
\begin{equation}
\left( \begin{array}{c}
a_{{\bf k}} \\ b^{\dagger}_{-{\bf k}} \\ b_{{\bf k}} \\
a^{\dagger}_{-{\bf k}} \end{array} \right) = S_{{\bf k}}
\left( \begin{array}{c}
\alpha_{{\bf k}} \\ \beta^{\dagger}_{-{\bf k}} \\ \beta_{{\bf k}} \\
\alpha^{\dagger}_{-{\bf k}} \end{array} \right)
\label{eq38}
\end{equation}
where the elements $s_{ij}$ of $S_{{\bf k}}$ obey
\begin{eqnarray}
\sum_{j} s^2_{ij}(-1)^{j+1} & = & (-1)^{i+1} \nonumber \\
\sum_{j} s_{ij}s_{kj}(-1)^{j+1} & = & 0 \nonumber \\
s_{ij} & = & s_{5-i,5-j}
\label{eq39}
\end{eqnarray}
Then
\begin{equation}
{\hat H}_{{\bf k}}' = S^T{\hat H}_{{\bf k}}S.
\label{eq40}
\end{equation}

There are initially 16 unknown parameters, corresponding to the
elements of the transformation matrix $S_{{\bf k}}$. The
conditions (\ref{eq39}) turn out to eliminate 12 of these, leaving
only 4 independent parameters; these 4 parameters are determined
by the condition that ${\hat H}_{{\bf k}}'$ should be diagonal,
i.e.
\begin{equation}
h_{12}' = h_{13}' = h_{14}' = h_{23}' = 0.
\label{eq41}
\end{equation}
We have determined the solution to this problem numerically.
In its diagonalized form, the Hamiltonian now reads:

\begin{eqnarray}
H_{{\bf k}} & = & 2S[h_{11}'(a^{\dagger}_{{\bf k}}\alpha_{{\bf k}} +
\alpha^{\dagger}_{-{\bf k}}\alpha_{-{\bf k}}) \nonumber \\
 & & +h_{22}' (\beta^{\dagger}_{{\bf k}}\beta_{{\bf k}}
+\beta^{\dagger}_{-{\bf k}}\beta_{-{\bf k}}) + N_{{\bf k}}'],
\label{eq42}
\end{eqnarray}
where
\begin{equation}
N_{{\bf k}}' = h_{11}' + h_{22}' -h_{11}-h_{22}
\label{eq43}
\end{equation}
Hence one can easily obtain numerical
results for the ground state energy and single-particle dispersion
relations.
The staggered magnetization is given by
\begin{eqnarray}
M_s & = & \frac{1}{N}
<\psi_0|\sum_BS^z_B + \cos \theta \sum_A S^z_A|\psi_0> \nonumber \\
 & = & \frac{S}{2}(1+\cos
\theta) \nonumber \\
 & & -\frac{2}{N}\sum_k[s^2_{21}+s^2_{23}+\cos \theta
(S^2_{12}+s^2_{14})]
\label{eq44}
\end{eqnarray}
and the net magnetization in the z direction:
\begin{eqnarray}
M_z & = & \frac{1}{N}
<\psi_0|\sum_BS^z_B - \cos \theta \sum_A S^z_A|\psi_0> \nonumber \\
 & = & \frac{S}{2}(1-\cos
\theta) \nonumber \\
 & & -\frac{2}{N}\sum_k[s^2_{21}+s^2_{23}-\cos \theta
(S^2_{12}+s^2_{14})]
\label{eq45}
\end{eqnarray}

\begin{table}
\caption{Ground-state energy per bond as a function of $\alpha$.
Linear spin-wave theory: $\epsilon_0^{(1)}$; modified second-order
spin-wave theory: $\epsilon_0^{(2)}$; with higher-order
corrections $\epsilon_0^{(3)}$.{\newline}}
\begin{tabular}{c|ccc|c}
\hline
$\alpha$ & & N{\' e}el expansion & & Canted expansion \\
 & $\epsilon_0^{(1)}$ & $\epsilon_0^{(2)}$ & $\epsilon_0^{(3)}$ & $\epsilon_0^{(1)}$ \\
\tableline
0.0 & -0.32897 & -0.33521 & -0.33503 & \\
0.1 & -0.31778 & -0.32523 & -0.32496 & \\
0.2 & -0.30666 & -0.31543 & -0.31509 & \\
0.3 & -0.29561 & -0.30586 & -0.30543 & \\
0.4 & -0.28465 & -0.29652 & -0.29602 & \\
0.5 & -0.27377 & -0.28746 & -0.28691 & -0.27377 \\
0.6 & -0.26300 & -0.27870 & -0.27817 & -0.25465 \\
0.7 & -0.25233 & -0.27028 & -0.26990 & -0.24458 \\
0.8 & -0.24180 & -0.26224 & -0.26224 & -0.24050 \\
0.9 & -0.23141 & -0.25462 & - & -0.24058 \\
1.0 & -0.22119 & -0.24746 & - & -0.24367 \\
\hline
\end{tabular}
\label{tab1}
\end{table}

\begin{table}
\caption{ Net magnetization $M_z$ and staggered magnetization
$M_s$ as functions of $\alpha$. Conventions as for Table
\ref{tab1}.{\newline}}
\begin{tabular}{c|ccc|cc}
\hline
$\alpha$ & & N{\' e}el expansion & & Canted & expansion\\
 & $M_s^{(1)}$ & $M_s^{(2)}$ & $M_s^{(3)}$& $M_s^{(1)}$ & $M_z^{1)}$ \\
\tableline
0.0 & 0.30340 & 0.30340 & 0.30601 & & \\
0.1 & 0.29582 & 0.29149 & 0.29499 & & \\
0.2 & 0.28759 & 0.27819 & 0.28265 & & \\
0.3 & 0.27859 & 0.26327 & 0.26872 & & \\
0.4 & 0.26873 & 0.24647 & 0.25270 & & \\
0.5 & 0.25785 & 0.22751 & 0.23386 & 0.25785 & 0.00000\\
0.6 & 0.24578 & 0.20607 & 0.21088 & 0.28800 & 0.02688\\
0.7 & 0.23230 & 0.18181 & 0.18163 & 0.30067 & 0.05143\\
0.8 & 0.21714 & 0.15442 & 0.14529 & 0.30525 & 0.07207\\
0.9 & 0.19992 & 0.12369 &    -    & 0.30579 & 0.08912\\
1.0 & 0.18017 & 0.08954 &    -    & 0.30429 & 0.10321\\
\hline
\end{tabular}
\label{tab2}
\end{table}

\section{Numerical Results}
\label{sec4}

Numerical results for the model have been obtained using the
finite-lattice method. The momentum sums were carried out for a fixed
sublattice size $L=\sqrt{N/2}$, using discrete values for the momenta
$k_x$ and $k_y$, e.g.

\begin{equation}
k_x(i) = \frac{2\pi(i-1/2)}{L}, \hspace{5mm} i=1, \cdots L,
\label{eq46}
\end{equation}
where we take half-integer values corresponding to anti-periodic
boundary conditions to avoid any integrable singularities at ${\bf k} =
0.$ Results were obtained for $L=8,9 \cdots 12$, and a fit in powers of
$1/L$ was made to extrapolate to the bulk limit $L \rightarrow \infty$.
The finite-size corrections for the ground-state energy per bond scale
asymptotically like $1/L^3$, and those for the magnetization like $1/L$
\cite{hasenfratz1993,zheng1993}. The resulting bulk estimates are shown in Tables \ref{tab1} and
\ref{tab2}. The values at $\alpha = 0$ agree with those obtained by
Gochev \cite{gochev1994} for the pure Heisenberg case.

\begin{figure}
 \includegraphics[width=1.0\linewidth]{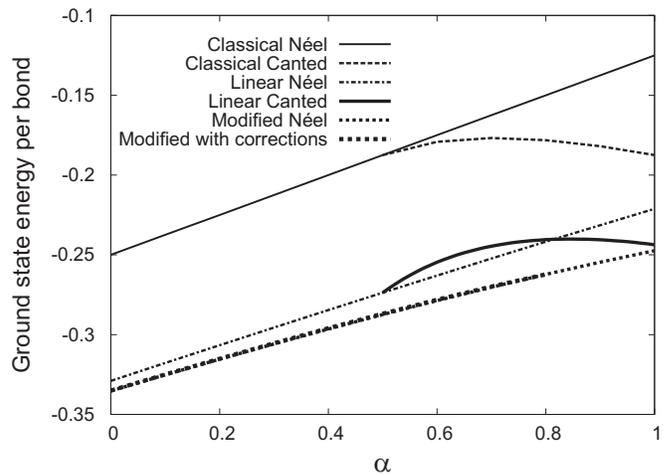}
 \caption{ Ground-state energy per bond as a function of coupling
$\alpha$. Short-dashed line: classical result; long-dashed line: linear
spin-wave theory; solid line: 2nd order spin-wave theory.}
 \label{fig6}
\end{figure}

Figure \ref{fig6} shows the behaviour of the
ground-state energy resulting from the classical calculation and linear
spin-wave theory for both the N{\' e}el phase and the canted phase,
and the modified treatment with corrections for the N{\'
e}el phase. In linear spin-wave theory,
the N{\' e}el and canted results coincide at $\alpha = 0.5$, as
they must do since $\theta = 0$ there. Somewhat surprisingly, however,
the N{\' e}el energy remains the lower of the two beyond that point,
until the two curves cross once more at $\alpha \simeq 0.84$. In other
words, the transition between the N{\' e}el and canted phases is
pushed out to $\alpha \simeq 0.84$ in linear spin-wave theory, and is
clearly first-order.
The modified treatment to second order lowers the energy a little
further, while the higher-order corrections are virtually negligible, and
indistinguishable on the diagram.

\begin{figure}
 \includegraphics[width=1.0\linewidth]{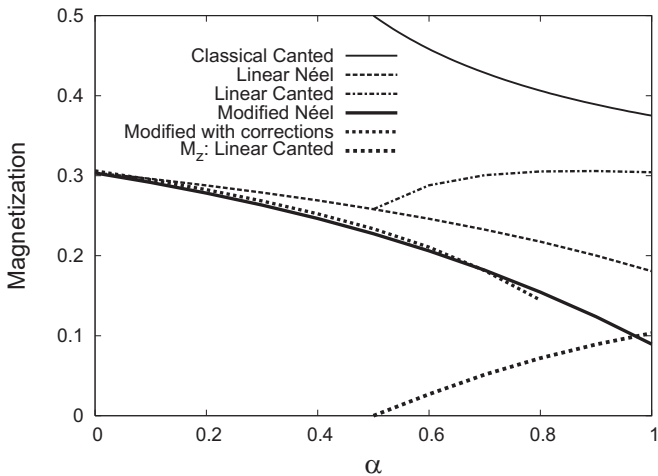}
 \caption{ Staggered magnetization $M_s$ and net magnetization $M_z$ as
functions of $\alpha$.}
 \label{fig7}
\end{figure}

Figure \ref{fig7} shows the magnetizations $M_s$ and $M_z$ as functions of the
coupling $\alpha$. In linear spin-wave theory, the staggered
magnetization $M_s$ in the N{\' e}el phase is reduced by quantum fluctuations,
as expected, but
the effective coupling is only half that in the
$J_1-J_2$
model, and so $M_s$ remains substantial at $\alpha
= 0.5$.
In the modified second-order treatment it is lowered somewhat further,
while the higher-order corrections are small and positive for low
$\alpha$, and turn negative beyond $\alpha \simeq 0.7$.
 The staggered magnetization in
the canted phase is also reduced by quantum fluctuations,
although the effect is reduced at large $\alpha$.

\begin{figure}
 \includegraphics[width=0.9\linewidth]{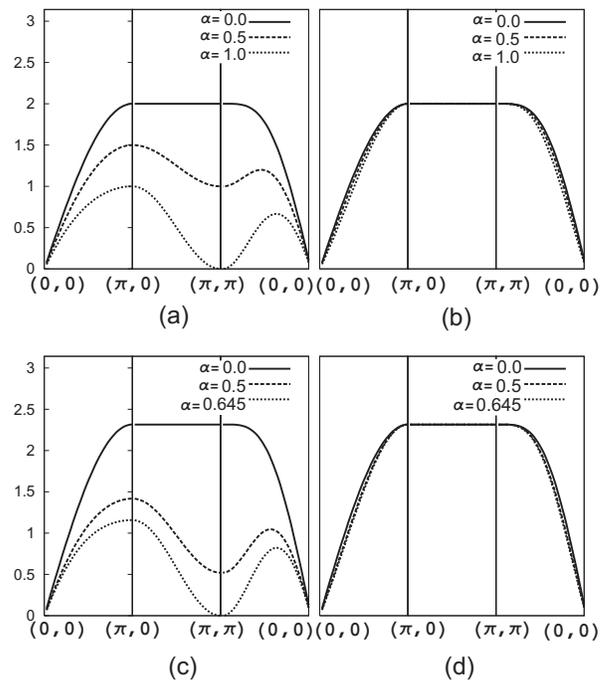}
 \caption{ Dispersion curves in the N{\' e}el phase along high-symmetry
lines in the Brillouin
zone for various $\alpha$.
Curves a) and b) are from linear spin-wave theory, c) and d) from the
modified second-order theory. Cases a) and c): $\alpha$ bosons; cases b)
and d): $\beta$ bosons.}
 \label{fig8}
\end{figure}

Figure \ref{fig8} illustrates the spin-wave dispersion of the $\alpha$
and $\beta$ bosons in the N{\' e}el phase as given by the second-order theory.
It can be seen that the dispersion curve for the $\beta$ bosons remains
virtually unchanged at all $\alpha$. That for the $\alpha$ bosons,
however, develops an instability at ${\bf k} = (\pi,\pi)$, and the
energy gap is predicted to vanish when $\alpha$ gets too large,
signalling a transition. The instability occurs at $\alpha=1.0$ in
linear spin wave theory, and $\alpha=0.645$ in the modified second-order
theory.

\begin{figure}
 \includegraphics[width=0.9\linewidth]{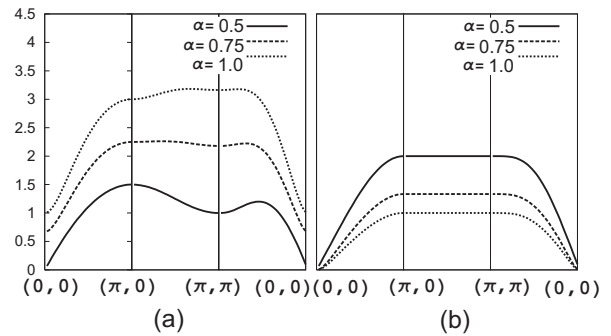}
 \caption{ Dispersion curves in the canted phase along
high-symmetry lines in the Brillouin
zone, obtained from linear spin-wave theory.
Case a): $\alpha$ bosons; b) $\beta$ bosons.}
 \label{fig9}
\end{figure}

Figure \ref{fig9} shows similar plots from the linear theory in
the canted phase. The energy gap for the $\beta$ bosons remains
zero at all couplings, with linear dispersion near the origin; but
as soon as we move away from $\alpha=0.5$ (i.e $\theta=0$) an
energy gap opens up for the $\alpha$ bosons, with quadratic
dispersion near the origin. This is in accordance with the general
counting of Goldstone modes in non-relativistic systems
\cite{nielsen1976}, and with previous discussions of Goldstone
modes in canted phases \cite{sachdev1996,dassarma1998,roman2000},
where the spontaneous symmetry breaking pattern is SU(2)
$\rightarrow Z_2$.

\section{Conclusions}
\label{sec5}

The Union Jack model is another example of a spin-1/2 Heisenberg
antiferromagnet with frustration on the square lattice. A classical
variational analysis predicts a transition at $\alpha = 0.5$ between the
N{\' e}el phase at small $\alpha$, and a canted ferrimagnetic phase at large
$\alpha$. Thus the system exhibits
a new phenomenon, namely ferrimagnetism induced by frustration.

In linear spin-wave theory, the transition is pushed out to $\alpha_c
\simeq 0.84$. This makes the model potentially much more interesting,
because it is very likely that the N{\' e}el magnetization will vanish
at or before that point. In the $J_1-J_2$ model, numerical studies
\cite{dagotto1989,schulz1996,kotov1999,gelfand1989,
singh1999,sushkov2001,capriotti2000,jongh2000}
show that the N{\' e}el magnetization vanishes at about
$\alpha_c \simeq 0.38$. The frustration in the Union Jack model is only
half as strong, as seen by a glance at Figure \ref{fig1} or from linear
spin-wave theory, and so one might expect the magnetization to vanish at
about $\alpha_c \simeq 0.76$ in this model.
 A plausible scenario, then, is that a second-order N{\' e}el
transition might occur at $\alpha_c \simeq 0.76$, possibly followed by
an intermediate spin-liquid phase as in the $J_1-J_2$ model, and then a
first-order transition to the canted phase at somewhat larger
$\alpha$. Numerical experiments would be necessary to ascertain if this
is indeed the case.

\acknowledgments
We would like to thank Profs. J. Oitmaa and O. Sushkov, and Dr. G.
Misguich, for very useful discussions and advice.
This work forms part of a research project supported by a grant
from the Australian Research Council.

\begin{widetext}
\appendix{}
\begin{center}
{\bf APPENDIX}
\end{center}

The vertex functions ${\Phi^{(i)},i=1 \cdots 9}$ are:

\begin{eqnarray}
\Phi^{(1)}({\bf1234}) & = &
\gamma(4-1)v_1u_2v_3u_4+\gamma(4-2)u_1v_2v_3u_4+\gamma(3-1)v_1u_2u_3v_4+\gamma(3-2)u_1v_2u_3v_4 \nonumber \\
& & -\gamma(4)u_1u_2v_3u_4-\gamma(3)u_1u_2u_3v_4-\gamma(3)v_1v_2v_3u_4-\gamma(4)v_1v_2u_3v_4\\
& & -\alpha Q u_1u_2v_3v_4 \nonumber \\
\Phi^{(2)}({\bf1234}) & = &
\gamma(4-2)v_1u_2u_3v_4+\gamma(3-2)v_1u_2v_3u_4+\gamma(4-1)u_1v_2u_3v_4+\gamma(3-1)u_1v_2v_3u_4 \nonumber \\
& & -\gamma(3)v_1v_2v_3u_4-\gamma(4)v_1v_2u_3v_4-\gamma(3)u_1u_2u_3v_4-\gamma(4)u_1u_2v_3u_4\\
& & -\alpha Q v_1v_2u_3u_4 \nonumber \\
\Phi^{(3)}({\bf1234}) & = &
\gamma(4-1)v_1u_2u_3u_4+\gamma(3-1)v_1u_2v_3v_4+\gamma(4-2)u_1v_2u_3u_4+\gamma(3-2)u_1v_2v_3v_4 \nonumber \\
& &-\gamma(4)u_1u_2u_3u_4-\gamma(3)u_1u_2v_3v_4-\gamma(3)v_1v_2u_3u_4-\gamma(4)v_1v_2v_3v_4\\
& &-\alpha Q u_1u_2u_3v_4 \nonumber \\
\Phi^{(4)}({\bf1234}) & = &
\gamma(3-1)u_1v_2u_3u_4+\gamma(4-1)u_1v_2v_3v_4+\gamma(3-2)v_1u_2u_3u_4+\gamma(4-2)v_1u_2v_3v_4 \nonumber \\
& &-\gamma(3)v_1v_2u_3u_4-\gamma(4)v_1v_2v_3v_4-\gamma(3)u_1u_2v_3v_4-\gamma(4)u_1u_2u_3u_4\\
& &-\alpha Q v_1v_2v_3u_4 \nonumber \\
\Phi^{(5)}({\bf1234}) & = &
\gamma(4-1)v_1v_2v_3u_4+\gamma(4-2)u_1u_2v_3u_4+\gamma(3-1)v_1v_2u_3v_4+\gamma(3-2)u_1u_2u_3v_4 \nonumber \\
& &-\gamma(4)u_1v_2v_3u_4-\gamma(3)u_1v_2u_3v_4-\gamma(3)v_1u_2v_3u_4-\gamma(4)v_1u_2u_3v_4\\
& &-\alpha Q u_1v_2v_3v_4 \nonumber \\
\Phi^{(6)}({\bf1234}) & = &
\gamma(4-1)u_1u_2u_3v_4+\gamma(4-2)v_1v_2u_3v_4+\gamma(3-1)u_1u_2v_3u_4+\gamma(3-2)v_1v_2v_3u_4 \nonumber \\
& &-\gamma(4)v_1u_2u_3v_4-\gamma(3)v_1u_2v_3u_4-\gamma(3)u_1v_2u_3v_4-\gamma(4)u_1v_2v_3u_4\\
& &-\alpha Q v_1u_2u_3u_4 \nonumber \\
\Phi^{(7)}({\bf1234}) & = &
\gamma(4-2)v_1u_2v_3u_4+\gamma(4-1)u_1v_2v_3u_4+\gamma(3-2)v_1u_2u_3v_4+\gamma(3-1)u_1v_2u_3v_4 \nonumber \\
& &-\gamma(4)v_1v_2v_3u_4-\gamma(3)v_1v_2u_3v_4-\gamma(3)u_1u_2v_3u_4-\gamma(4)u_1u_2u_3v_4\\
& &-\alpha Q v_1v_2v_3v_4 \nonumber \\
\Phi^{(8)}({\bf1234}) & = &
\gamma(4-2)u_1v_2u_3v_4+\gamma(4-1)v_1u_2u_3v_4+\gamma(3-2)u_1v_2v_3u_4+\gamma(3-1)v_1u_2v_3u_4 \nonumber \\
& &-\gamma(4)u_1u_2u_3v_4-\gamma(3)u_1u_2v_3u_4-\gamma(3)v_1v_2u_3v_4-\gamma(4)v_1v_2v_3u_4\\
& &-\alpha Q u_1u_2u_3u_4 \nonumber \\
\Phi^{(9)}({\bf1234}) & = &
\gamma(4-2)u_1u_2u_3u_4+\gamma(4-1)v_1v_2u_3u_4+\gamma(3-2)u_1u_2v_3v_4+\gamma(3-1)v_1v_2v_3v_4 \nonumber \\
& &-\gamma(4)u_1v_2u_3u_4-\gamma(3)u_1v_2v_3v_4-\gamma(3)v_1u_2u_3u_4-\gamma(4)v_1u_2v_3v_4\\
& &-\alpha Q v_1u_2v_3u_4 \nonumber
\end{eqnarray}
where
\begin{equation}
Q  = \eta(3-2)+\eta(4-2)-\eta(3)-\eta(4).
\end{equation}
\end{widetext}


\begin{references}
\bibitem{anderson1987} P.W. Anderson, Science (Washington, DC, U.S.) {\bf
235}, 1196 (1987).
\bibitem{chandra1988} P. Chandra and B. Doucot, Phys. Rev. B {\bf 38}, 9335 (1988).
\bibitem{dagotto1989} E. Dagotto and A. Moreo, Phys. Rev. Lett. {\bf
63}, 2148 (1989); F. Figueirido {\it et al.}, Phys. Rev. B{\bf
41}, 4619 (1990).
\bibitem{schulz1996} H.J. Schulz, T.A.L. Ziman and D. Poilblanc, J.
Phys. I {\bf 6}, 675 (1996).
\bibitem{kotov1999} V.N. Kotov, J. Oitmaa, O.P. Sushkov and Zheng W-H., Phys. Rev. B {\bf
60}, 14613 (1999).
\bibitem{read1991} N. Read and S. Sachdev, Phys. Rev. Lett. {\bf 66}, 1773 (1991); {\it
ibid} {\bf 62}, 1694 (1989); G. Murthy and S. Sachdev, Nucl. Phys. B{\bf
344}, 557 (1990).
\bibitem{gelfand1989} M.P. Gelfand, R.R.P. Singh and D.A. Huse, Phys.
Rev. B {\bf 40}, 10801 (1989); M.P. Gelfand, {\it ibid}, {\bf 42}, 8206
(1990).
\bibitem{singh1999} R.R.P. Singh, Zheng W-H., C.J. Hamer and J. Oitmaa,
Phys. Rev. B {\bf 60}, 7278 (1999).
\bibitem{sushkov2001} O.P. Sushkov, J. Oitmaa and Zheng W-H., Phys. Rev.
B {\bf 63}, 104420 (2001).
\bibitem{capriotti2000} L. Capriotti and S. Sorella, Phys. Rev. Lett.
{\bf 84}, 3173 (2000).
\bibitem{sorella2003} S. Sorella, L. Capriotti, F. Becca and A. Parola,
Phys. Rev. Lett. {\bf 91}, 257005 (2003).
\bibitem{capriotti2003} L. Capriotti, F. Becca, A. Parola and S. Sorella,
Phys. Rev. B{\bf 67}, 212402 (2003).
\bibitem{jongh2000} M.S.L. du Croo de Jongh, J.M.J. van Leeuwen and W.
van Saarloos, Phys. Rev. B {\bf 62}, 14844 (2000).
\bibitem{yamanaka1997} M. Yamanaka, M. Oshikawa and I. Affleck, Phys.
Rev. Lett. {\bf 79}, 1110 (1997).
\bibitem{oshikawa2000} M. Oshikawa, Phys. Rev. Lett. {\bf 84}, 1535
(2000).
\bibitem{hastings2004} M.B. Hastings, Phys. Rev. B{\bf 69}, 104431
(2004).
\bibitem{takano2003} K. Takano, Y. Kito, Y. Ono and K. Sano, Phys. Rev.
Lett. {\bf 91}, 197202 (2003).
\bibitem{misguich2002} G. Misguich, C. Lhuillier, M. Mambrini and P.
Sindzingre, Eur. Phys. J. B{\bf 26}, 167 (2002).
\bibitem{gochev1994} I.G. Gochev, Phys. Rev. B{\bf49}, 9594 (1994).
\bibitem{xu1990} J.H. Xu and C.S. Ting, Phys. Rev. B {\bf 42}, 6861
(1990).
\bibitem{nishimori1990} H. Nishimori and Y. Saika, J. Phys. Soc. Jpn.
{\bf 59}, 4454 (1990); A.F. Baranov and O.A. Starykh, JETP Lett. {\bf
51}, 311 (1990); T. Oguchi and H. Kitakani, J. Phys. Soc. Jpn. {\bf 59},
3322 (1990).
\bibitem{hirsch1989} J.E. Hirsch and S. Tang, Phys. Rev. B {\bf 39},
2887 (1989).
\bibitem{takahashi1987} M. Takahashi, Phys. Rev. Lett. {\bf 58}, 168
(1987); Phys. Rev. B{\bf 40}, 2494 (1989).
\bibitem{hasenfratz1993} P. Hasenfratz and F. Niedermayer, Z. Phys.
B{\bf 92}, 91 (1993).
\bibitem{zheng1993} Zheng W-H. and C.J. Hamer, Phys. Rev. B{\bf 47},
7961 (1993).
\bibitem{nielsen1976} H.B. Nielsen and S. Chada, Nucl. Phys. B{\bf 105},
445 (1976).
\bibitem{sachdev1996} S. Sachdev and T. Senthil, Ann. Phys. (N.Y.) {\bf
251}, 76 (1996).
\bibitem{dassarma1998} S. Das Sarma, S. Sachdev,and L. Zheng, Phys. Rev.
B{\bf 58}, 4672 (1998).
\bibitem{roman2000} J.M. Roman and J. Soto, Phys. Rev. B{\bf 62}, 3300
(2000).






\end{references}
\end{document}